# Phase-referenced Nonlinear Spectroscopy of the α-Quartz/Water Interface

Paul E. Ohno,[1] Sarah A. Saslow,[1,#] Hong-fei Wang,[2] Franz M. Geiger,[1]*, and Kenneth B. Eisenthal[3]

[1]Department of Chemistry, Northwestern University, Evanston, IL 60208, [#]now Earth Systems Science Division, Energy & Environment Directorate, Pacific Northwest National Laboratory, Richland, WA 99352, [2]Physical Sciences Division, Physical & Computational Sciences Directorate, Pacific Northwest National Laboratory, Richland, WA 99352, [3]Department of Chemistry, Columbia University, New York, NY 10027, USA

*Corresponding author: geigerf@chem.northwestern.edu

Probing the polarization of water molecules at charged interfaces by second harmonic generation spectroscopy has been heretofore limited to isotropic materials. Here, we report non-resonant nonlinear optical measurements at the interface of anisotropic z-cut α-quartz and water under conditions of dynamically changing ionic strength and bulk solution pH. We find that the product of the third-order susceptibility and the interfacial potential, $\chi^{(3)} \cdot \Phi(0)$, is given by $(\chi^{(3)} - i\chi^{(3)}) \cdot \Phi(0)$, and that the interference between this product and the second-order susceptibility of bulk quartz depends on the rotation angle of α-quartz around the z-axis. Our experiments show that this newly identified term, $i\chi^{(3)} \cdot \Phi(0)$, which is out of phase from the surface terms, is of bulk origin. The possibility of internally phase referencing the interfacial response for the interfacial orientation analysis of species or materials in contact with α-quartz is discussed along with the implications for conditions of resonance enhancement.

**Introduction.** The strength of the second harmonic electric field, $E_{2\omega}$, that is produced at charged interfaces is a function of the incident fundamental electric field, $E_\omega$, the second-order susceptibility of the interface, $\chi^{(2)}$, the zero-frequency electric field corresponding to the interfacial potential produced by surface charges, $\Phi(0)$, and the third-order susceptibility, $\chi^{(3)}$, according to[1-6]

$$\sqrt{I_{SHG}} \propto E_{2\omega} \propto \chi^{(2)} E_\omega E_\omega + \chi^{(3)} E_\omega E_\omega \Phi(0) \qquad (1)$$

Early work, in which the relative phase of the terms contributing to the SHG intensity was included[7], shows that when the wavelength of the fundamental and second harmonic photons are far from electronic and vibrational resonance, $\chi^{(2)}$ and $\chi^{(3)}$ are real, though they may differ in sign[7,8]. Yet, phase information has not been recovered in traditional second harmonic generation (SHG) detection schemes, as they only collect the square modulus of the signal. While phase information from SHG and vibrational sum frequency generation (SFG) signals can be readily obtained through coherent interference of the signal of interest with an external[9-16] or internal[17,18] phase standard, applications of such reference techniques to determine the phase of SHG signals generated at buried interfaces, such as charged oxide/water interfaces, is challenging due to the presence of dispersive media on both sides of the interface. Additionally, the interface between water and α-quartz, the most abundant silicate mineral in nature[19-21], has been theoretically predicted to produce a more ordered interfacial water layer than amorphous silica[22-24], though this has not yet been probed using even traditionally detected SHG, as the non-centrosymmetric bulk generally produces second harmonic signals that overpower surface SHG signals by orders of magnitude to the point where the surface signal is indistinguishable from the bulk response. Doing so under non-resonant conditions, however, would circumvent interaction terms between possibly potential-dependent $\chi^{(2)}$ and $\chi^{(3)}$ contributions to which resonantly enhanced SHG[25] or SFG[16] studies of charged interfaces may be sensitive.

Here, we present an experimental geometry that produces considerable non-resonant SHG signal intensity from the z-cut α-quartz/water interface in the presence of bulk SHG signals from both the quartz and the electrical double layer under conditions of dynamically varying pH and ionic strength (Figure 1). The approach, which uses an external reflection geometry, femtosecond laser pulses having just nanojoule pulse energies, and a high repetition rate, enables us to experimentally identify a source of surface potential-induced bulk SHG from the electrical double layer. Further, it expands the scope of SHG spectroscopy to probe interfaces of non-centrosymmetric materials and establishes phase-referenced SHG spectroscopy to buried interfaces by using z-cut α-quartz as an internal phase standard.

**Results.**

**pH Jumps over Silica and Quartz/Water Interfaces.** Using our previously described dual-pump flow system[26], we transition the pH of the aqueous phase between pH 3 and 11.5 so as to probe the interfacial potential dependence of the SHG responses. Near the point of zero charge, reported variously in the literature as pH 2.2[27] and pH 2.6[28] for α-quartz and pH 2.3 for fused silica[29], little SHG signal from the interface is expected, whereas the considerable negative interfacial potential at pH 11.5 should yield considerable SHG signal intensity[1,26]. Figure 2 shows the SHG vs time traces obtained from the fused silica/water interface using the flow cell shown in Figure 1. Indeed, we observe the same increases (respectively decreases) in SHG signal intensity upon increasing (respectively decreasing) the bulk solution pH that we previously reported for fused silica using a total internal reflection geometry (see Supplementary Figure 1)[26]. At 3 mM total salt concentration, SHG signal intensities generally range between 5 and 10 counts per second at pH 3, and between 30 and 40 counts per second at pH 11.5. Upon replacing the fused silica window with right-handed z-cut α-quartz oriented 30° from the +x-axis (see Methods), we observe the same response, albeit with a large bulk signal intensity leading to both significantly larger overall signal intensities and variations in signal intensity with varying pH.

**Rotating Quartz Orientation Angle Reveals Interference.** Figure 2 shows that the change in the SHG intensity observed for high vs. low pH flips sign upon rotation of the quartz crystal by 60° around the z-axis, indicating modulation between negative and positive interference with the $\chi^{(2)}$ term of the bulk quartz. The P-in/P-out polarization combination (termed "PP" hereafter) was selected as it demonstrated the highest interfacial sensitivity out of the PP, PS, 45P, and 45S polarization combinations surveyed (see Supplementary Figure 2). Figure 3 further shows that the constructive and destructive interference of the SHG signal depends on changing crystal rotation angles for the PP polarization combination; Supplementary Figure 3 shows the $\Delta I_{SHG}$ as a function of quartz rotational angle. Because in this geometry the beams must propagate through the dynamically changing aqueous phase, the dependence of the observed changes in the SHG intensity on the rotational angle of the α-quartz substrate indicates their origin as the interface and not changing optical properties of the aqueous phase, which would not depend on the angle of the α-quartz substrate. Additional control experiments show invariance of the results with minor variations in focal lens position

(Supplementary Figure 4) and quadratic dependence of the $I_{SHG}$ on laser power (Supplementary Figure 5), as expected.

**Discussion.** The constructive and destructive interference seen in Figures 2 and 3 can be understood by recalling that the non-resonant SHG or SFG signal from bulk α-quartz is a purely imaginary term[30,31]. This property of α-quartz has been employed to provide an internal phase standard that can amplify and interfere with the imaginary part of the vibrational SFG spectra of molecular surface species[17,18]. Yet, when the surface second-order susceptibility is non-resonant, i.e. all the surface response terms are purely real, interference with the imaginary term of the bulk α-quartz response cannot occur. However, the observations shown in Figure 2 demonstrate that the non-resonant SHG signal from the water/α-quartz interface is subject to interference from the non-resonant SHG signal from the bulk α-quartz, indicating a new source of surface potential-induced bulk SHG from the aqueous solution.

The observed interference shown in Figure 2 can be rationalized by considering that the phase of the bulk SHG signal produced by α-quartz shifts by 180° when the crystal is rotated clockwise by 60° around the z-axis (the sign of the bulk quartz phase is depicted in Figure 1 for our geometry). Rotation is shown to change the coherent interaction between the interfacial signal and bulk signal from constructive to destructive interference, as is expected from a 180° phase shift in the bulk quartz $\chi^{(2)}$ term[17]. This behavior can be mathematically treated according to the following equation:

$$I_{\text{SHG}} \propto \left| \chi^{(2)} + (\chi_1^{(3)} - i\chi_2^{(3)})\Phi(0) \pm i\chi_{\text{bulk quartz}}^{(2)} \right|^2 \tag{2}$$

where the sign of the $\pm i\chi^{(2)}_{\text{bulk quartz}}$ term is controlled by the rotational angle of the α-quartz crystal. Here, $\chi^{(3)}_1$ and $i\chi^{(3)}_2$ are related by a phase matching factor as described in Supplementary Note 1. Even though the $\chi^{(3)}$ mechanism for interfacial potential-induced second harmonic generation has been long established[1], the importance of phase matching in the $\chi^{(3)}$ term has only recently been theoretically considered[32]. An experimental validation requires a phase-referenced measurement, like the one demonstrated and established herein.

As further confirmation that the observed changes in the SHG signal intensity with pH are attributable to the interface, we recorded SHG signal intensities at pH 7 under conditions of increasing ionic strength for two quartz crystal rotation angles differing by 60°. For a charged interface, increases in the ionic strength result in screening of the interfacial charges, thus reducing the interfacial potential to which the water molecules in the electrical double layer are subjected, and ultimately the associated SHG signal intensity[5,33]. Indeed, this behavior is observed for both fused silica (Supplementary Figure 6) and α-quartz (Figure 4), at ionic strengths above ~$10^{-4}$ M NaCl, with opposite behavior (increases in the ionic strength coincide with SHG intensity increases) observed upon rotating the crystallographic axis of the z-cut quartz by 60°, as is observed in the pH jump experiments. We thus confirm the constructive and destructive interference discovered here for the α-quartz/water interface.

Our findings greatly expand the scope of SHG spectroscopy beyond amorphous and centrosymmetric materials and towards crystal classes that lack centrosymmetry, including the more than 500 non-centrosymmetric oxides catalogued to date[34]. In doing so, they open a path for directly comparing the amphoteric properties of amorphous and crystalline materials, such as fused silica and α-quartz. The implications of using an SHG bulk signal as an internal reference from which phase information of the surface signal – and therefore orientation information at the interface – can be determined are intriguing. The use of thin film deposition techniques such as atomic layer deposition, electron beam deposition, or spin-coating, or surface functionalization methods such as silane chemistry, directly on α-quartz, or other reference materials with known phase and second-order susceptibilities, offers the option of using phase referenced SHG spectroscopy as a method for unambiguously determining the sign (+ or -) of surface charges. A thin film with a thickness of only a few nm would be expected to interact with water molecules in a similar fashion to the surface of its bulk material, yet still allow coherent interaction between the second harmonic signal generated at the interface and the bulk signal produced by the α-quartz substrate. Likewise, a thin film of a nonlinear optical crystal with a known phase grown on fused silica or another optically transparent centrosymmetric medium would allow for the presence of an internal phase standard with an acceptable SHG intensity without requiring propagation through an aqueous medium. Yet, for conditions of electronic or vibrational resonance, we caution that the absorptive (imaginary) and dispersive (real) terms of $\chi^{(2)}$, $\chi_1^{(3)}$ and $\chi_2^{(3)}$ may mix.

**Methods.**

**Sample Information.** In the experiments, we worked with three different right-handed, z-cut α-quartz samples (10 mm diameter, 3 mm thick) from three different vendors: Meller Optics (Providence, RI); Knight Optical (North Kingston, RI); and Precision Micro-Optics (Woburn, MA). The fused silica sample was purchased from Meller Optics. Prior to measurements, the samples were treated for 1 hour with NoChromix solution

(Godax Laboratories), a commercial glass cleaner (*caution: NoChromix should only be used after having read and understood the relevant safety information*). The samples were then sonicated in methanol for six minutes, sonicated in Millipore water for six minutes, dried in a 100°C oven, and plasma cleaned (Harrick Plasma) for 30 seconds on the highest setting. This procedure produces surfaces that are void of vibrational SFG responses in the C-H stretching region[26].

**Determining Crystal Orientation.** Due to the dependence of the $I_{SHG}$ response on the orientational angle of the α-quartz crystal sample, it was necessary to unambiguously identify the crystal orientation used in the experiments. In this study, we define $\varphi$ to be the clockwise rotation of the crystal about its z-axis, measured from its +x-axis (*i.e.* at 0° the incoming laser beam is aligned with its horizontal projection along the +x-axis of the α-quartz crystal, at 30° the crystal has been rotated 30° clockwise, etc). The x-axis of the crystal can be determined by finding the $I_{SHG}$ maximum in the PP polarization combination or the $I_{SHG}$ minimum in the PS polarization combination[17] (Supplementary Figure 7) while rotating the crystal about its axis. Determining the orientation of the x-axis, *i.e.* whether the incoming laser beam is aligned parallel or anti-parallel with the x-axis, is more difficult. Possible techniques include measuring the sign of the small voltage produced upon deformation of the crystal due to its piezoelectricity,[35] determining whether the bulk signal constructively or destructively interferes with the SFG C-H stretching signal from alkane chain monolayers absorbed on the interface,[17] or obtaining Laue diffraction patterns from the α-quartz crystal. We compared Laue diffraction patterns from an α-quartz crystal of known orientation (provided by the supplier) with that of our unknown sample in order to determine its absolute orientation

(Supplementary Figure 8). We also obtained two commercial α-quartz samples for which the suppliers (Knight Optical and PM Optics) had determined and provided us the absolute crystal orientation and obtained the same SHG responses across all three samples (Supplementary Figure 3B).

**Laser Setup.** We focused the p-polarized 800 nm output of a Ti:Sapphire oscillator (Mai Tai, Spectra-Physics, 100 fs pulses, λ=800 nm, 82 MHz) through a hollow fused quartz dome onto the interface between water and the solid substrate in the external reflection geometry depicted in Figure 1. Prior to the sample stage, the beam was passed through a long-pass filter to remove any second harmonic co-propagating with the fundamental beam, attenuated with a variable density filter to 0.50 ± 0.01 W, passed through a half-wave plate for input polarization control, and focused onto the α-quartz/water or fused silica/water interface at an angle of 60°. The beam waist in the focal region is estimated at 30 $\mu$m in diameter.

The SHG signal was recollimated, passed through a 400 nm bandpass filter (FBH400-40, Thor Labs), and directed through a polarizer and monochromator for detection via a Hamamatsu photomultiplier tube connected to a preamplifier (SR445A, Stanford Research Systems) and a single photon counter (SR400, Stanford Research Systems), as detailed in our earlier work[26]. A portion of the fundamental beam was picked off prior to the sample stage and continuously monitored by a power meter (Newport 1917-R) during acquisitions to allow for continuous normalization of signal intensity to power and account for the impact of slight, albeit unavoidable, laser power fluctuations on the SHG signal intensity. Compared to internal reflection, SHG signals obtained using the present geometry are generally ~400 times more sensitive to the interface relative to the bulk (see Supplementary Table 1).

We note that geometries in which 100-femtosecond pulses from a kHz amplifier laser system accessed the quartz/water interface through bulk quartz as thin as 200 μm (i.e. the inverted geometry of what is depicted in Figure 1a) were not successful, even when applying a second quartz plate for background suppression[36].

**Solution Preparation.** The aqueous solutions were prepared using Millipore water (18.2 MΩ/cm) and NaCl (Alfa Aesar, 99+%). The pH of the solutions was adjusted using dilute solutions of HCl (E.M.D., ACS grade) and NaOH (Sigma Aldrich, 99.99%) and verified using a pH meter.

**Data Availability.** All relevant data are available from the authors upon request to the corresponding author.

**Acknowledgments.** Signal sensitivity analysis and measurement performed by PEO, SAS, and FMG was supported by the U.S. National Science Foundation (NSF) under graduate fellowship research program (GRFP) awards to PEO and SAS, and award number CHE-1464916 to FMG. Method design/development by FMG and SAS was supported by the U.S. Department of Energy, Office of Science, Office of Basic Energy Sciences, Chemical Sciences, Geosciences, and Biosciences Division. HFW was supported by the Materials Synthesis and Simulation Across Scales (MS3) Initiative through the LDRD program at Pacific Northwest National Laboratory (PNNL). PNNL is multi-program national laboratory operated for Department of Energy by Battelle under Contracts No. DE-AC05-76RL01830. KBE gratefully acknowledges NSF award number CHE-1057483.

**Author Contributions.** FMG, HFW, and KBE conceived of the idea. PEO and SAS performed the experiments. FMG, HFW, and PEO analyzed the data. The manuscript was written with substantial contributions from FMG, HFW, PEO, and SAS.

**Author Information.** The authors declare no competing financial interests. Correspondence should be addressed to FMG (geigerf@chem.northwestern.edu).

**Figure Captions.**

**Figure 1 | Experimental setup.** (**a**) Depiction of the external reflection geometry and the flow cell enabling in this work. FL=focal lens, RL=recollimating lens. (**b**) Top view of right-handed z-cut α-quartz as placed across the plane of incidence, and depiction of the sign of the bulk quartz phase. (**c**) Depiction of the phase-referencing process used to unveil the newly identified term $i\chi^{(3)}\cdot\Phi(0)$.

**Figure 2 | pH Jump experiments.** SHG intensity in counts per second obtained from the fused silica/water (gray trace, left y-axis) and α-quartz/water (blue and red traces, right y-axis) interfaces under conditions of dynamically changing bulk solution pH varying between 3 and 11.5. Blue and red trace indicate results obtained with 60° difference in α-quartz rotational angle. The constructive and destructive interference observed for the α-quartz/water interface confirm interference between the potential-dependent interfacial and bulk quartz terms of the nonlinear susceptibility.

**Figure 3 | Interference experiment.** SHG intensity vs time traces normalized to intensity at pH 3 obtained from the α-quartz/water interface at different rotational angles of the α-quartz crystal during conditions of dynamically changing bulk solution pH varying between 3 and 11.5.

**Figure 4. | Salt Screening experiment.** SHG intensity from the α-quartz/water interface maintained at pH 7 during conditions of dynamically changing bulk solution ionic strength varying between $10^{-5}$ and $10^{-1}$ M NaCl. Error bars represent one standard deviation, with six (respectively nine) measurements for the φ=75° (respectively 15°) series.

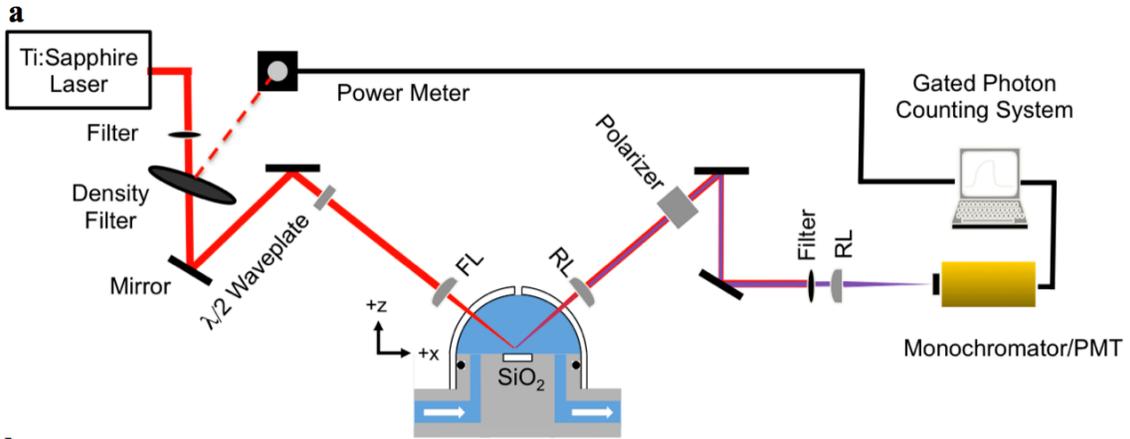
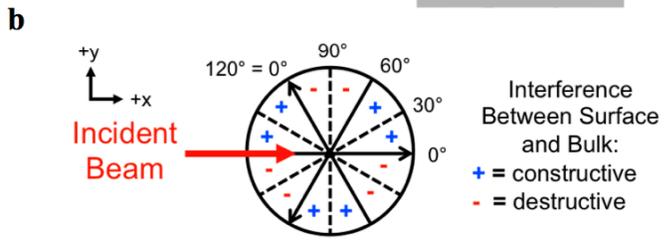
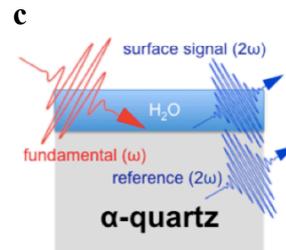

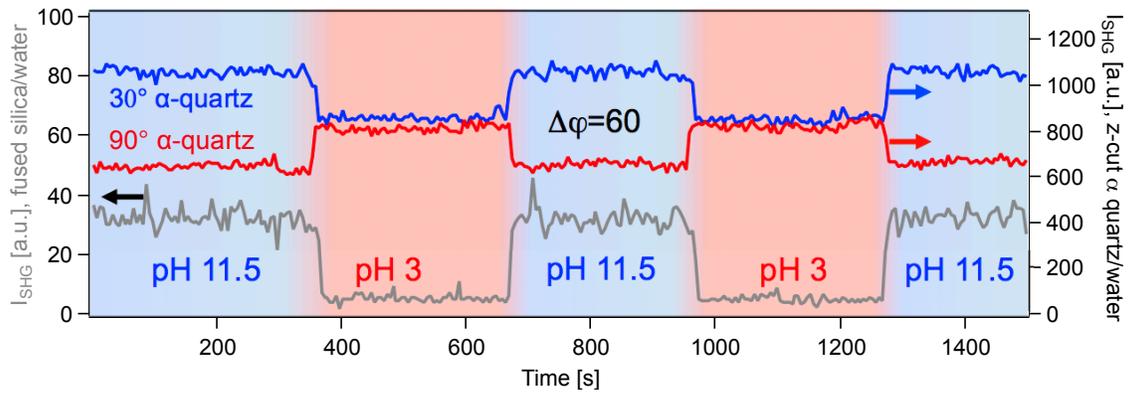

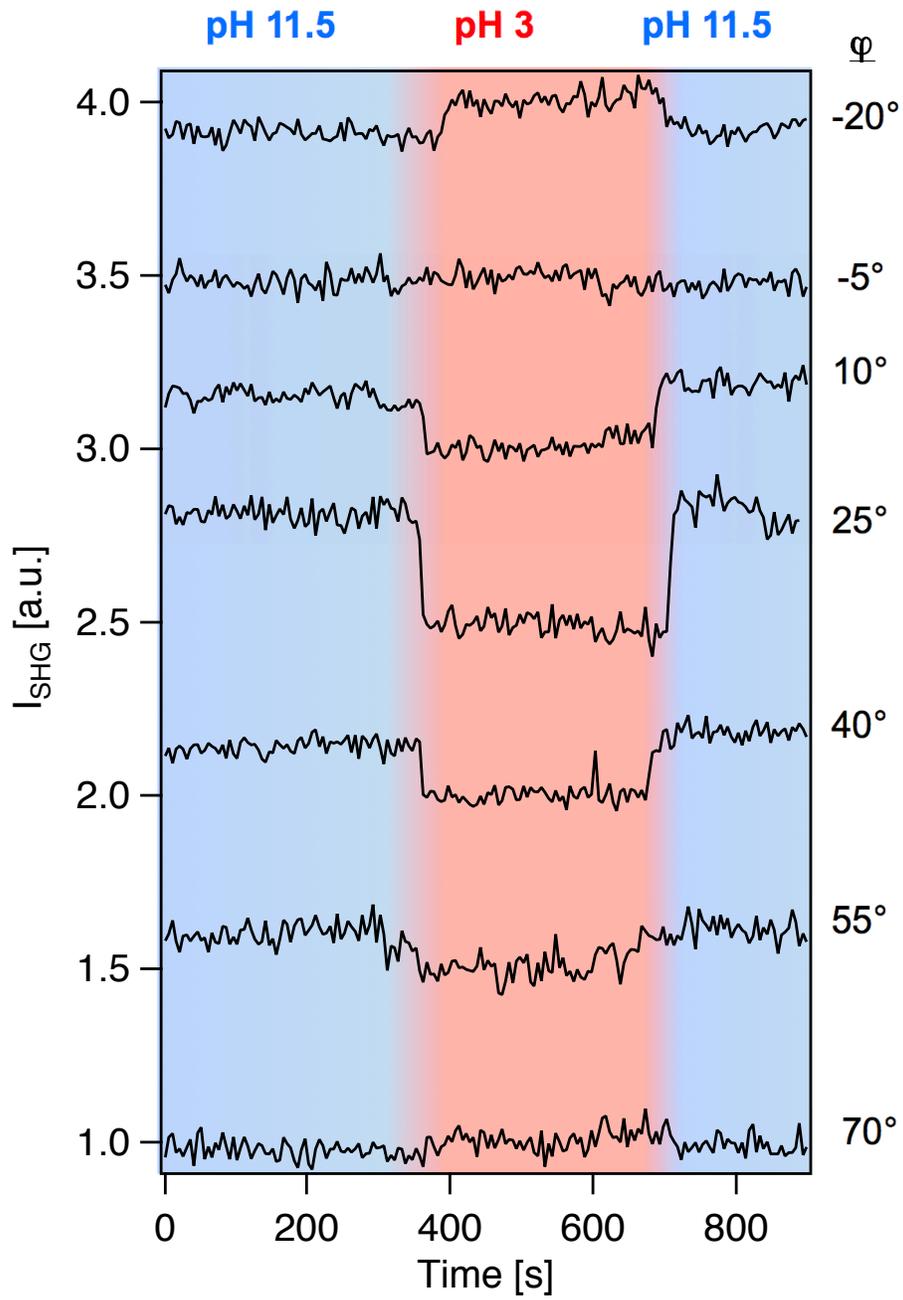

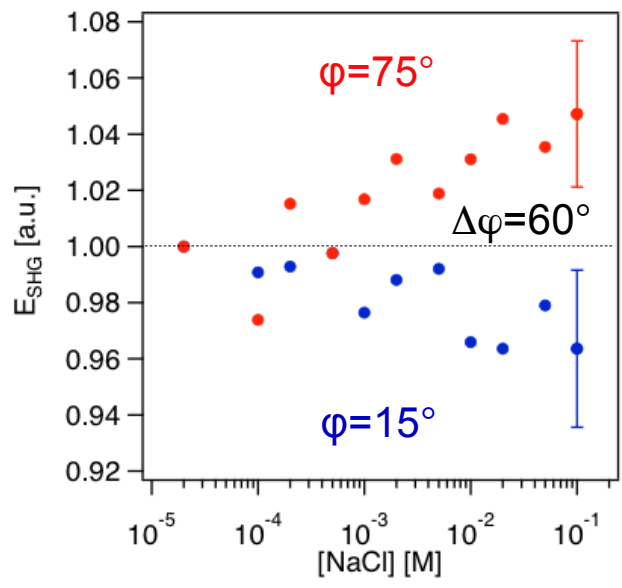

**Supplementary Information.**

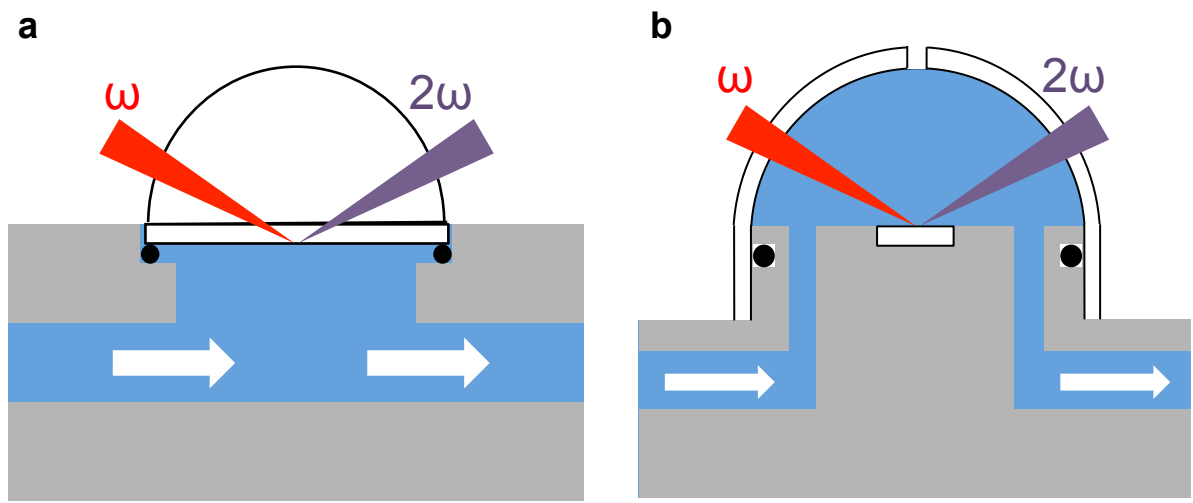

**Supplementary Figure 1 |** (**a**) Schematic depictions of the total-internal or near total-internal reflection geometry. (**b**) Schematic depictions of the external reflection geometry. Internal reflection has been used generally by our lab[1,2] and others[3] as it produces a higher signal intensity due to higher Fresnel coefficients.[4] We utilize fused silica hemispheres in order to ensure high transmission coefficients and retain the input polarization at any input angle. To study surfaces besides fused silica in this geometry, disks of material such as muscovite[5] can be clamped between the hemisphere and the aqueous phase. However, the non-centrosymmetric nature of α-quartz produced an overwhelmingly large bulk signal in this geometry, motivating the switch to the external reflection geometry (see Table S1).

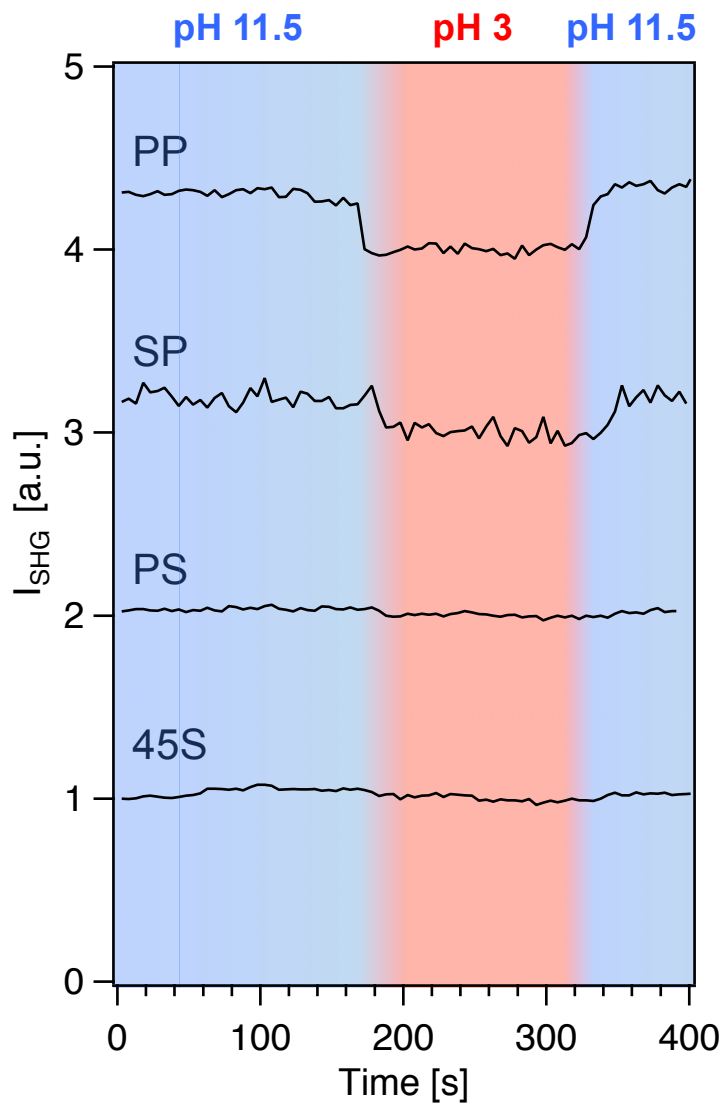

**Supplementary Figure 2** | Polarization combination dependence of the α-quartz $I_{SHG}$ response to pH jump experiments (the first index represents input polarization, the second represents output polarization; 45 represents mixed polarization). The crystal was oriented at 30° and the traces are normalized to their intensities at low pH; each is offset by 1 unit. The PP polarization combination, showing the highest sensitivity to interfacial potential, was used in all other experimental results reported unless otherwise specified.

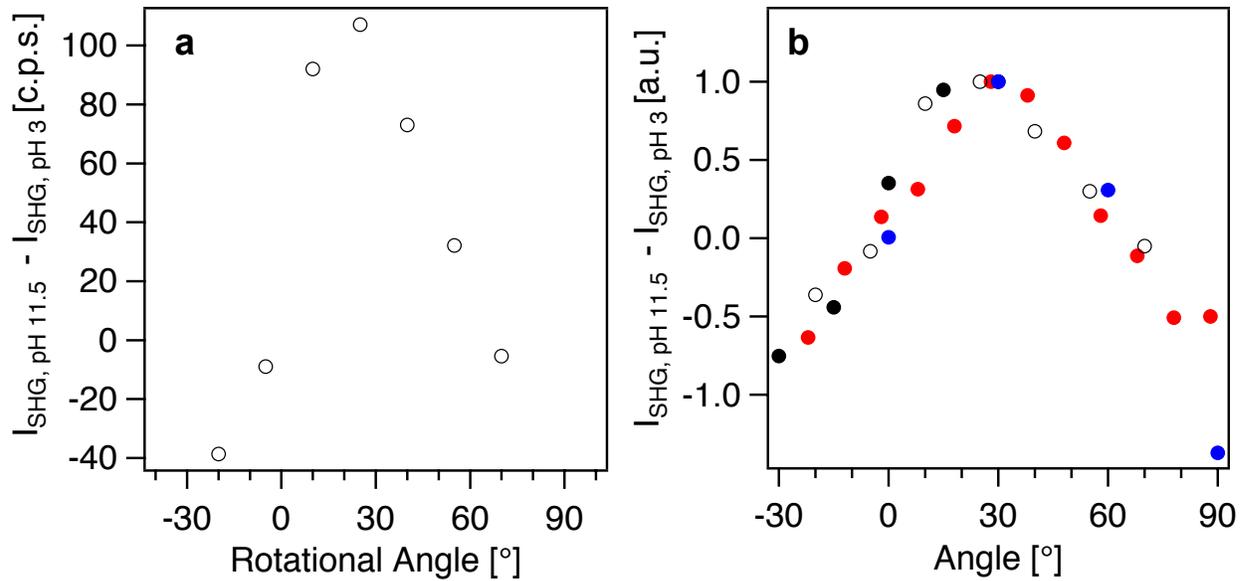

**Supplementary Figure 3 | (a)** The difference in $I_{SHG}$ at low and high pH conditions, $\Delta I_{SHG}$, as a function of rotational angle of the α-quartz crystal for the data set included in Figure 3 in the main text. **(b)** $\Delta I_{SHG}$, as a function of rotational angle of the α-quartz crystal for all experiments carried out on three different samples from different suppliers, normalized to the maximum difference obtained in each run (blue circles, PM Optics; open black circles, Meller Optics Experimental Run 1; closed black circles, Meller Optics Experimental Run 2; red circles, Knight Optical). The same overall pattern holds accross of the samples of little to no change at 0° and 60°, a maximum of constructive interference at 30°, and a maximum of destructive interference at -30° (= 90°).

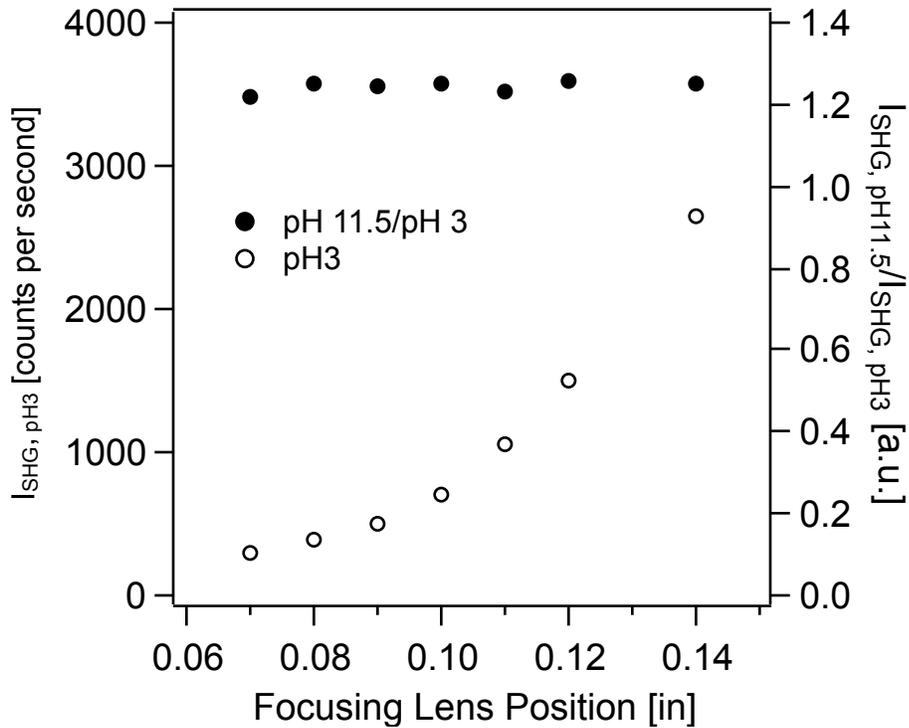

**Supplementary Figure 4** | Signal intensity dependence upon focusing lens position. While the signal intensity at pH 3 (open circles, left axis), which is predominately from the bulk $\chi^{(2)}$ term of the α-quartz, is dependent upon the focusing lens position, the normalized jump intensity given by the signal at pH 11.5 divided by the signal at pH 3 (filled circles, right axis) is constant across a range of focusing lens positions that produce pH 3 signal intensities from ~300 to ~2600 counts per second. This allows attribution of changes in these normalized jump intensities across different rotational angles to the angles themselves as opposed to small changes in focusing lens position caused by physical rotation of an imperfectly aligned sample stage.

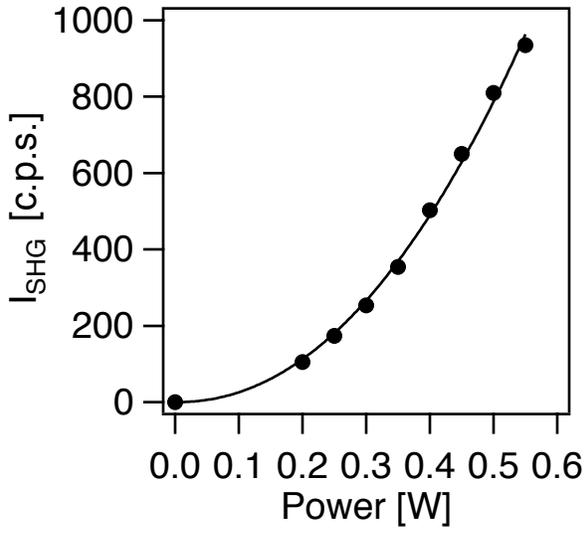

**Supplementary Figure 5** | The power dependence of $I_{SHG}$ for the α-quartz/water interface at pH 3; exponent in power fit = 2.1(1) as expected for a second order process such as SHG.

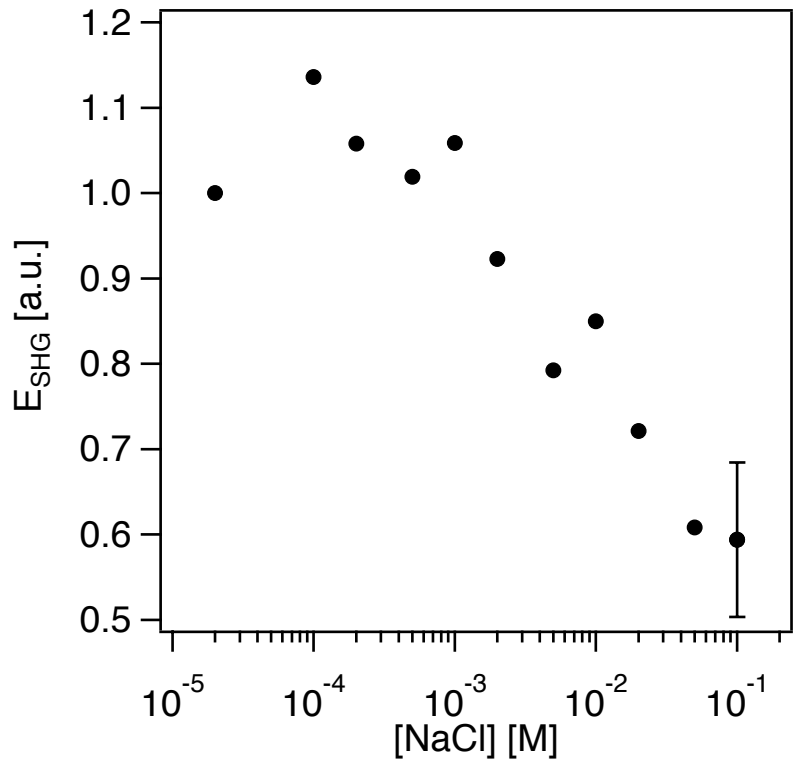

**Supplementary Figure 6** | SHG intensity from the fused silica/water interface maintained at pH 7 during conditions of dynamically changing bulk solution ionic strength varying between $10^{-5}$ and $10^{-1}$ M NaCl. Interfacial potential and charge screening properties of the fused silica/water interface are expected to be comparable to those of the α-quartz/water interface, though the SHG signal lacks the bulk $\chi^{(2)}$ of the α-quartz due to the centrosymmetric nature of fused silica. This allows direct comparison to Figure 4 of the main text. Error bars represent one standard deviation with three measurements.

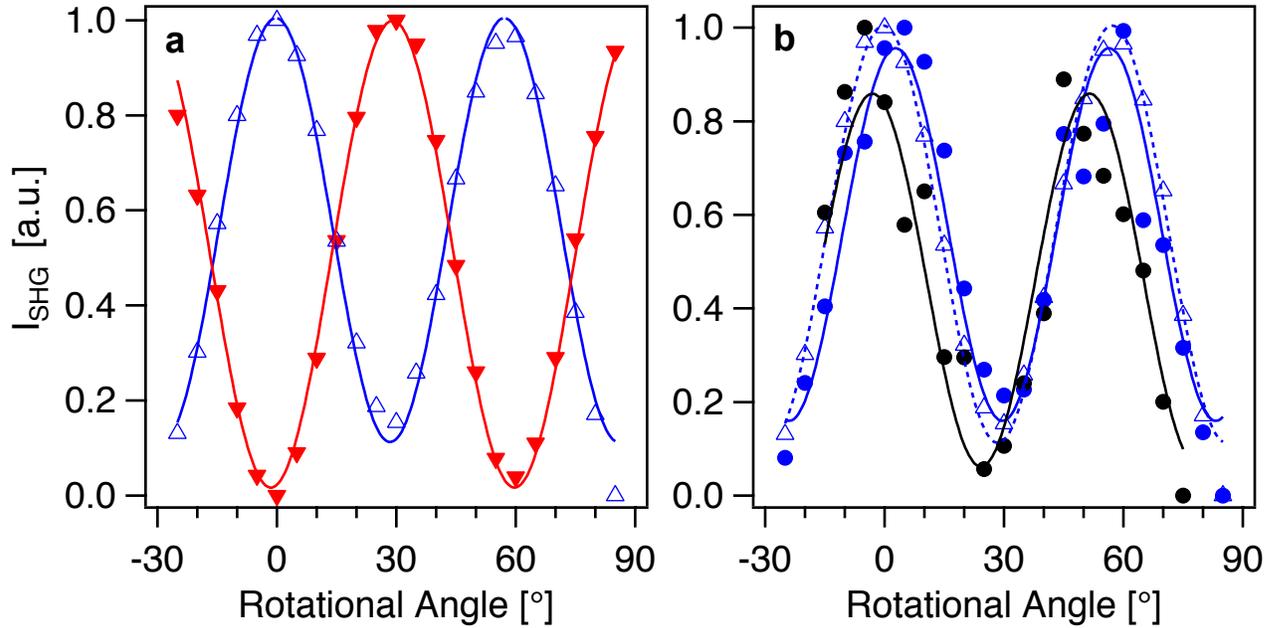

**Supplementary Figure 7 |** (**a**) Rotational angle dependence of $I_{SHG}$ for PP (blue, open triangles) and PS (red, upside down closed triangles) of bulk α-quartz measured in air using the reflection geometry described in the Supplementary Methods section. (**b**) Rotational angle dependence of $I_{SHG}$ in the PP polarization combination measured in air (blue, open triangles), with the hollow fused silica dome but no water present (blue, filled circles), and with the fused silica dome and pH 3 water present (black, filled circles), and using the reflection geometry described in the Supplementary Methods section. Imperfections in the spherical nature of the hand-blown dome introduces some error to the rotational dependence of the signal, but the periodic pattern is still present when going from the quartz/air (A) to the quartz/water (B) interface.

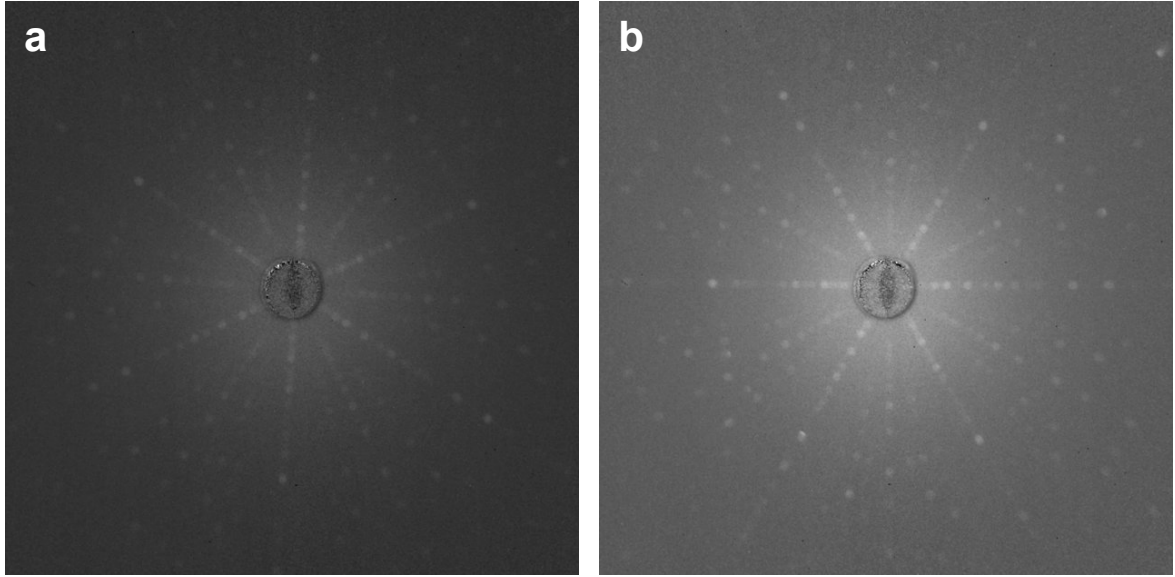

**Supplementary Figure 8 | (a)** Laue diffraction pattern of α-quartz samples of previously known absolute orientation (specimen from Knight Optical). **(b)** Laue diffraction pattern of α-quartz sample (specimen from Meller Optics) of unknown absolute orientation, used to identify the absolute orientation of the unknown orientation α-quartz crystal through direct comparison.

**Supplementary Table 1** | Typical signal intensities measured using the internal and external reflection geometries schematically depicted in Supplementary Figure 1B. The internal reflection geometry gives higher signal intensities from both the bulk α-quartz and the fused silica surface at pH 11.5, as expected from the Fresnel factors.[4] However, if it is assumed that the fused silica and α-quartz surfaces have roughly comparable signal intensities at pH 11.5, it can be seen that the ratio of expected surface signal intensity to bulk signal intensity for α-quartz is more favorable in the external reflection geometry, motivating its experimental use in this study.

|  | Typical $I_{SHG}$ [c.p.s.] | |
|---|---|---|
|  | Internal Reflection | External Reflection |
| α-Quartz | 1,500,000 | 1,000 |
| Fused Silica | 200 | 50 |
| **Ratio** | **7,500** | **20** |

**Supplementary Note 1**

**A brief derivation of the origin of the $i\chi_2^{(3)}$ term, see also Reference 32 in the Main Text.**

As the electric field $E_{dc}(z) = -d\Phi(z)/dz$ is z (depth)-dependent and there is the phase matching factor that is also z dependent, one has:

$$\begin{aligned}\chi_{dc}^{(2)} &= \int_0^\infty \chi^{(3)} E_{dc}(z) e^{-i\Delta k_z z} dz \\ &= \int_0^\infty -\chi^{(3)} \frac{d\Phi(z)}{dz} e^{-i\Delta k_z z} dz \\ &= -\chi^{(3)} \Phi(z) e^{-i\Delta k_z z} \Big|_0^\infty + \chi^{(3)} \int_0^\infty \Phi(z)(-i\Delta k_z) e^{-i\Delta k_z z} dz \\ &= \chi^{(3)}\Phi(0) + i\Delta k_z \chi^{(3)} \int_0^\infty \Phi(z) e^{-i\Delta k_z z} dz\end{aligned} \quad (1)$$

Here, $1/\Delta k_z$ is the coherence length of the SHG or SFG process, $\Phi(\infty) = 0$, and the following integration relationship was used:

$$\int \frac{df(z)}{dz} g(z) dz = f(z)g(z) - \int f(z) \frac{dg(z)}{dz} dz \quad (2)$$

A good approximation is that $\Phi(z) = \Phi(0) e^{-kz}$, where $1/k$ is the Debye screening length factor. Then,

$$\begin{aligned}\chi_{dc}^{(2)} &= \chi^{(3)}\Phi(0) + i\Delta k_z \chi^{(3)} \int_0^\infty \Phi(0) e^{-kz} e^{-i\Delta k_z z} dz \\ &= \chi^{(3)}\Phi(0) + \frac{-i\Delta k_z}{k + i\Delta k_z} \chi^{(3)}\Phi(0) \\ &= \frac{k}{k + i\Delta k_z} \chi^{(3)}\Phi(0)\end{aligned}$$

Therefore, in the total effective surface susceptibility,

$$\chi_{eff}^{(2)} = \chi^{(2)} + \chi_{dc}^{(2)} = \chi^{(2)} + (\chi_1^{(3)} - i\chi_2^{(3)})\Phi(0) \quad (3)$$

one has

$$\chi_1^{(3)} = \frac{k^2}{k^2 + (\Delta k_z)^2} \chi^{(3)} \qquad (4.1)$$

$$\chi_2^{(3)} = \frac{k \Delta k_z}{k^2 + (\Delta k_z)^2} \chi^{(3)} \qquad (4.2)$$

Therefore, because the surface field is real and the phase matching factor is complex, the total $\chi_{dc}^{(2)} = (\chi_1^{(3)} - i\chi_2^{(3)})\Phi(0)$ contribution is complex.

When $k \ll \Delta k_z$, i.e. the Debye length is long (low electrolyte concentration), one finds

$$\chi_1^{(3)} \sim 0 \text{ and } \chi_2^{(3)} \sim \frac{k}{\Delta k_z} \chi^{(3)} \qquad (5.1)$$

and the *dc* contribution is essentially imaginary.

When $k \gg \Delta k_z$, i.e. the Debye length is very small (high electrolyte concentration), one finds

$$\chi_1^{(3)} \sim \chi^{(3)} \text{ and } \chi_2^{(3)} \sim \frac{\Delta k_z}{k} \chi^{(3)} \sim 0 \qquad (5.2)$$

and the real term dominates.

When $k \sim \Delta k_z$, i.e. the Debye length and phase matching coherent length are comparable, the real and imaginary terms for the $\chi^{(3)}$ are comparable.

The derivation above assumes that the surface potential is of the form $\Phi(z) = \Phi(0) e^{-kz}$. The actual surface potential may be different from this form, but essentially it decays when moving away from the surface. In addition, the surface potential can not only induce bulk $\chi^{(3)}$ responses from the water side, but also from the fused silica or the α-quartz side.[6] These issues warrant further investigation in the future.

Nevertheless, the following relationship, as established herein, should generally hold:

$$\chi_{eff}^{(2)} = \chi^{(2)} + \chi_{dc}^{(2)} = \chi^{(2)} + (\chi_1^{(3)} - i\chi_2^{(3)})\Phi(0) \qquad (6)$$

**Supplementary References**